\documentclass[12pt, preprint]{aastex}
\newcommand\ea{et al.}
\newcommand\dg{$^\circ$}

\shorttitle{Solar meridional flows}
\shortauthors{Basu \& Antia}

\begin{document}

\title{Characteristics of solar meridional flows during solar cycle 23}

\author{Sarbani Basu}
\affil{Department of Astronomy, Yale University, P. O. Box 208101, New
Haven, CT 06520-8101; sarbani.basu@yale.edu}

\author{H. M. Antia}
\affil{Tata Institute of Fundamental Research, Homi Bhabha Road, Mumbai 400005, India;
antia@tifr.res.in}

\begin{abstract}

We have analyzed available full-disc data from the Michelson Doppler Imager (MDI)
on board {\it SoHO} using the ``ring diagram'' technique to determine the behavior
of solar meridional flows over solar cycle 23 in the outer 2\% of the solar
radius. 
We find that the dominant component of meridional flows during solar maximum was much lower than that
during the minima at the beginning of cycles 23 and 24. 
There were differences in the flow velocities even between the 
two minima.
The meridional flows show a migrating pattern with higher-velocity flows 
migrating towards the equator as activity increases.
Additionally, we find that the migrating pattern of the meridional flow
matches those of sunspot butterfly diagram and the zonal flows in the
shallow layers.
A high latitude band in meridional flow appears around
2004, well before the current activity minimum.
A Legendre polynomial decomposition of the meridional
flows shows that the latitudinal pattern of the
flow was also different during the maximum as compared to that during the two minima.
The different components of the flow have different time-dependences,
and the dependence is different at different depths.

\end{abstract}


\section{Introduction}
\label{sec:intro}

Meridional flow plays a key role in the operation of solar
dynamo.  Even as early as the 1970s, these flows were  recognized as important in explaining the
solar dynamo  and differential rotation (e.g., Durney 1975; Bogart \& Gierasch 1979).
It is believed to be responsible for transporting magnetic
elements towards the poles (Wang et al.~1989).
It also plays a role in generation of differential rotation and its
temporal variations (Glatzmaier \& Gilman 1982; Spruit 2003).
The meridional flows are also invoked as a means to
transport flux towards the equator near the base of the convection zone in some solar dynamo models (e.g., Choudhuri
et al.~1995; Dikpati \& Charbonneau 1999). There are
no measurements of the meridional flow in the deeper 
layers of the Sun and hence, near-surface measurements, along with other assumptions,  can be  used
to constrain solar dynamo theories.

There is a rich history of the study of the meridional
component of flows on the solar surface.
(e.g., Schr\"oter \& W\"ohl 1975; Howard 1979; Topka et al.~1982; Ribes et al.~1985).
However, because the  meridional-flow speed (about 20 m s$^{-1}$) is 
more than an order of magnitude smaller than other flows at the solar
surface, the early observations did not give a clear picture of these flows.
Thanks to the improvements in observational techniques,
the meridional flow has been reliably measured at the solar surface
using both magnetic elements as tracers (Komm et al.~1993; Hathaway \& Rightmire 2010)
and Doppler velocities (e.g., Hathaway 1996). These observations show a general
flow towards the poles in both hemispheres that is roughly antisymmetric
about the equator. 

The development of local helioseismic techniques
enabled the study of meridional flow inside the Sun.
Using the time-distance technique Giles et al.~(1997) showed that meridional
flow persists in the subsurface layers. Schou \& Bogart (1998), Basu et al.~(1999),
Haber et al.~(2000) and others applied the
ring diagram technique to study the meridional flow in the subsurface
layers up to a depth of about 15 Mm below the surface. The local helioseismic
techniques used in these studies do not give any information about
the deeper layers. In particular, these could not study the layers close
to the base of the solar convection zone which is believed to be the
region of the return flow from the poles to the equator.
On the other hand, global modes that  can
give information about deeper layers are rather insensitive to meridional
flow (Roth \& Stix 2008; Chatterjee \& Antia 2009) and hence are
not useful for this purpose. Nevertheless, even the behavior
of meridional flow in the immediate sub-surface layers can  give useful information
to constrain  dynamo models. Some models of zonal flows (Spruit 2003)
predict a steep variation with depth in the meridional flow near the surface
and hence these models too can be tested with the available seismic data.

The availability of seismic data over a full solar cycle has made it
also possible to study the time variations of the meridional flows.
There have been such studies with data that cover only a part of the
solar cycle
(Chou \& Dai 2001; Haber et al.~2002; Basu \& Antia 2003; Gonz\'alez Hern\'andez et al.~2006, 2008, 2010;
Zaatri et al.~2006; etc.). All these studies have found significant temporal variations
in the meridional flows, though they are not always in complete agreement with
each other (see Antia \& Basu 2007). 
Among the results that stand out is that the meridional flow speed is 
slower at maximum activity than at the minimum (Chou \& Dai~2001; subsequently
confirmed by Hathaway \& Rightmire~2010).
Additionally the  pattern
of the meridional flow also shows bands of fast or slow flows similar to
well known zonal flows (Beck et al.~2002; Gonz\'alez
Hern\'andez et al.~2010).

In this work we use data  from the Michelson Doppler Imager (MDI) on
board {\it SoHO} to study how the sub-surface meridional flows
changed over solar cycle 23.
We use the ring diagram technique
to study the temporal, latitudinal and depth variations in meridional flow.
We chose to use MDI data since it samples the
entire solar cycle and hence, may give a clear picture of the temporal
variations. 
It may be noted that although MDI instrument has been functioning through
the entire cycle 23, except for a short break around 1998, the ring diagram
data are available only during the so-called ``dynamics runs'', which typically
cover only 2-3 months every year, earlier during the
cycle and for shorter periods since 2003 when {\it SoHo}'s high-gain
antenna started malfunctioning. Thus although the entire solar cycle is
sampled by the ring diagram data, the coverage is not continuous and the
fill factor is rather low.
Data from the Global Oscillation Network Group (GONG) are
available for the declining phase of cycle 23 only. These have been
analyzed by Gonz\'alez Hern\'andez et al.~(2010). GONG data has the
advantage of continuous coverage in time from July 2001 onwards.
MDI magnetic field data have been used by
Hathaway \& Rightmire (2010) to study surface meridional flows.

\section{Data}
\label{sec:data}

The basic data set consists of `ring diagrams' obtained from
full-disk Dopplergrams from the MDI instrument.
Ring diagrams (Hill 1988) are three-dimensional (3D) power spectra of short-wavelength
modes in a small region of the Sun. High-degree (short-wavelength) modes
can be approximated as plane waves over a small area of the Sun as long as
the horizontal wavelength of the modes is much smaller than the solar
radius. Ring diagrams are obtained from a time series of Dopplergrams of
a specific area of the Sun tracked at the mean rotation velocity.
 A detailed description of the ring diagram technique
is given by Patr\'on \ea\ (1997),  Basu \ea\ (1999) and
Antia \& Basu (2007). Ring diagram analysis
has the advantage that, unlike global-mode analysis, it can be used to study
the horizontal components of solar flows as a function of latitude, longitude,
depth and time. In the present work, we average the spectra over longitudes
and study the variation of meridional flow as a function of latitude, depth
and time.

We use publicly available MDI ring diagrams for this work. Each ring
diagram was obtained from a time series of 1664 images covering 16\dg\ in longitude and latitude. 
These square regions are then apodized to circular regions of diameter $15^\circ$.
Successive spectra are separated by 15\dg\ in heliographic longitude of the central meridian.
These data are available for fifteen latitudes from 52.5\dg S to 
52.5\dg N in steps of 7.5\dg. Wherever possible, we have averaged 
power spectra for each latitude over one full Carrington rotation. This 
has not always been possible since MDI operates in the dynamics mode only
for a small fraction of the time. These incomplete sets were chosen to get
a reasonable sampling of the different phases of cycle 23.
Averaging over a number of spectra improves the statistics and reduces the
error estimates in the inferred velocity components.
Characteristics of the
sets used are listed in Table~\ref{tab:data}. Also listed in the table is the
10.7 cm radio flux as an indicator of the solar activity over the period
covered by the data.
The 10.7 cm radio flux is obtained from the NOAA, National Geophysical
Data Center\footnote{%
ftp://ftp.ngdc.noaa.gov/STP/SOLAR\_DATA/SOLAR\_RADIO/FLUX/Penticton\_Adjusted/daily/DAI\-LYPLT.ADJ}.
Results obtained from some of these data sets had
been presented by Basu \& Antia~(2003), however, we reanalyzed those
sets in order to be consistent in our analysis. 

Each ring diagram is fitted with the model described by Basu \& Antia (1999).
The parameter that describes the frequency shift due to meridional flow was then inverted 
using the methods of Optimally Localized Averages (OLA)
and Regularized Least Squares (RLS) to determine the variation of meridional velocity
with depth. The inversion results are generally
reliable in the depth range of about 1 to 14 Mm ($r=0.98R_\odot$ to $0.999 R_\odot$) as seen by the averaging
kernels (Antia \& Basu 2007), and the match between OLA and RLS results.

\section{Results}
\label{sec:res}

\begin{figure*}
\epsscale{0.85}
\plotone{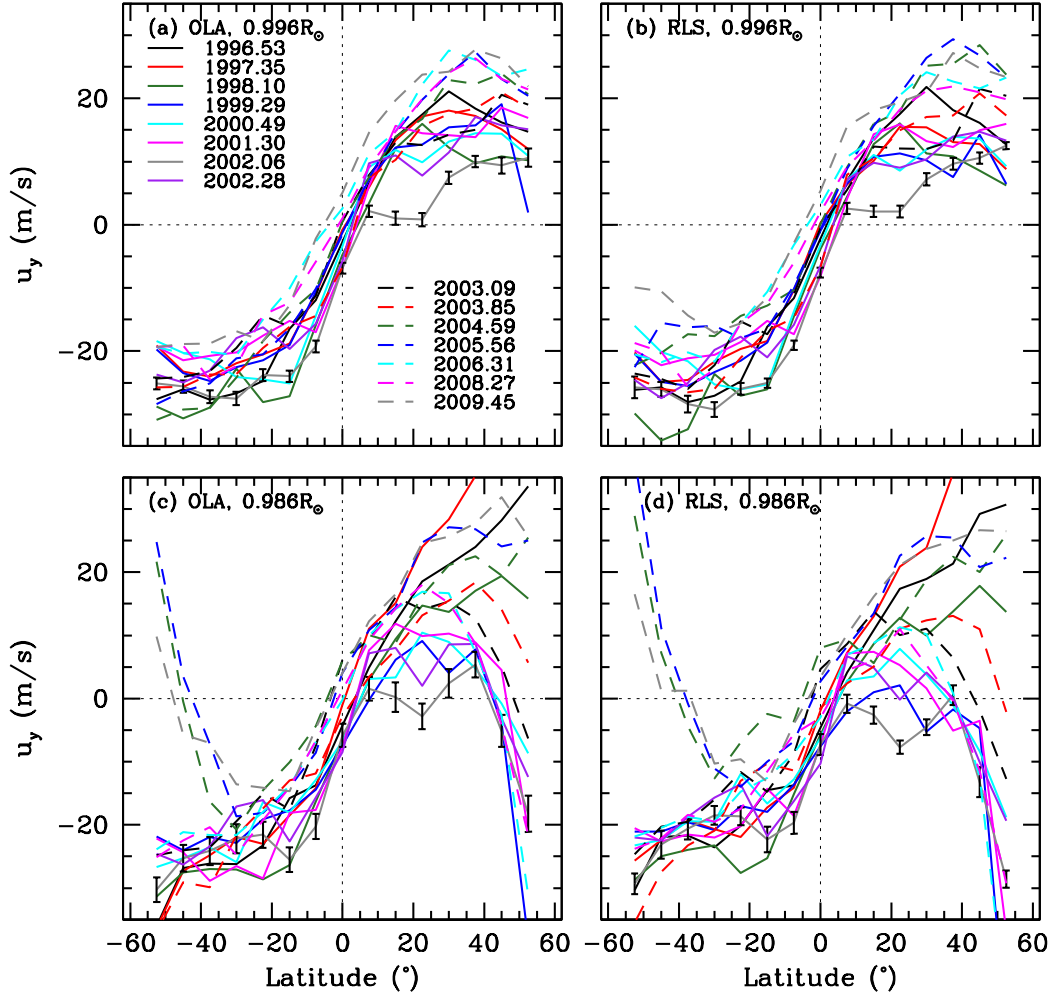}
\caption{The solar meridional flow velocities as obtained from the both OLA and RLS inversions
at depths of 2.8 Mm ($r=0.996R_\odot$) and 9.8 Mm ($r=0.986R_\odot$) are  plotted as a function of latitude.
The curves are labeled by the central time for each data set.
Positive values
indicate northward velocities and negative indicate southward velocities. For the
sake of clarity, error bars
are plotted on only one curve.}
\label{fig:full}
\end{figure*}

The meridional flow velocities at depths of 2.8 Mm ($r=0.996R_\odot$) and 9.8 Mm
($r=0.986R_\odot$)  obtained from our analysis 
are shown in Figure~\ref{fig:full}. We show both OLA and RLS results. Note that 
both results are by and large similar, giving added confidence in the results.
In some cases (e.g., 2009.45 at $r=0.996R_\odot$) there are some differences between RLS
and OLA results at high latitudes, but these are still within about $3\sigma$ of
each other. Considering the large number of results shown in the figure a few
differences of this order are to be expected. Furthermore, most of the large differences
are at high latitudes where projection effects become important.
This effect is not included in the errorbars which in turn results in an underestimation
of errors at high latitudes. Thus in real terms, the significance of the differences
are  even less than what they appear to be from Figure~\ref{fig:full}.
There is a clear temporal variation in the meridional flow velocities.
While the dominant flow is from the equator towards the poles in
both hemispheres, in some cases there appears to be a flow across the equator.
Such flows can be an artifact of position-angle errors (see
e.g., Giles et al.~1997), and hence, we ignore these in our analysis and
concentrate only on the component that is antisymmetric about the
solar equator. The antisymmetric component is obtained by taking an average over both
hemispheres, with sign of velocities in the south hemisphere reversed,
i.e. $(u_N-u_S)/2$.

\begin{figure}
\epsscale{0.9}
\plotone{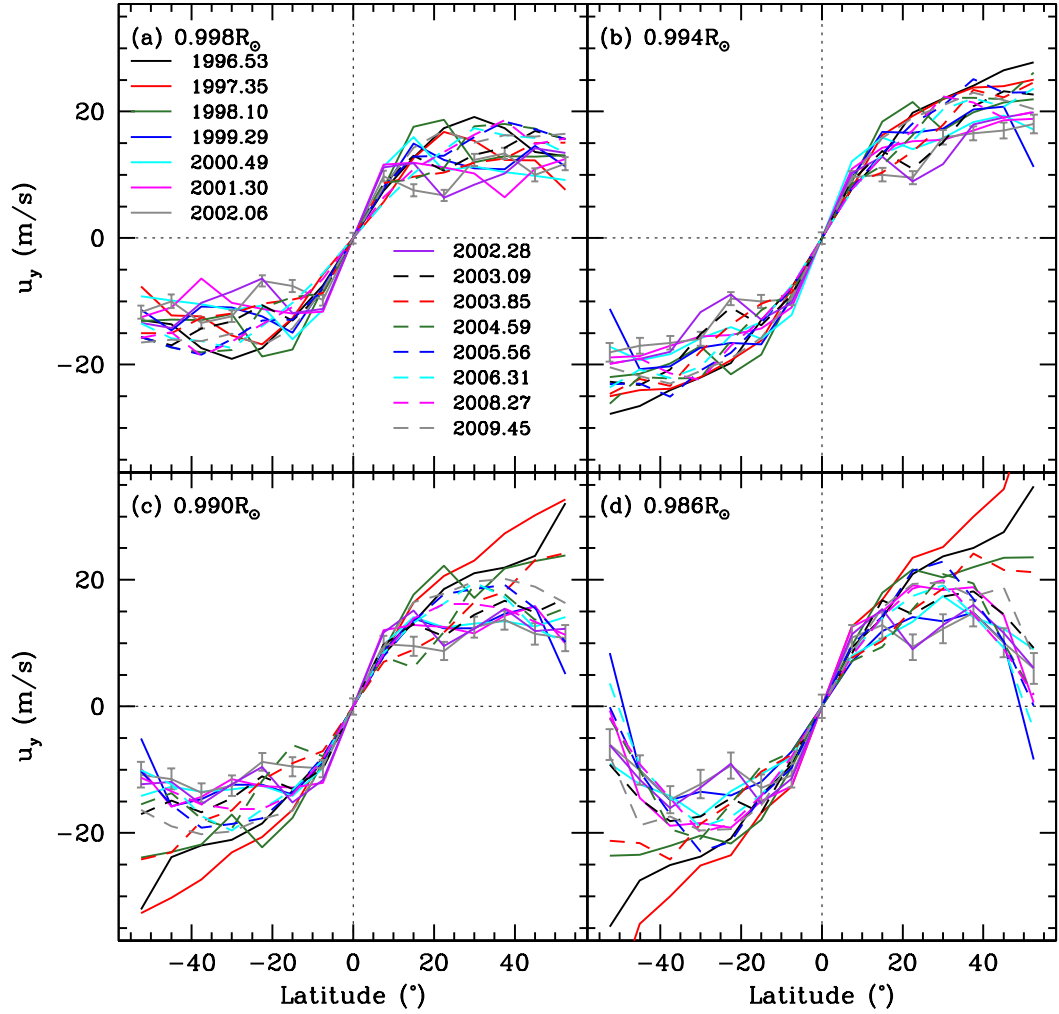}
\caption{The velocity of the  north-south antisymmetric component of the meridional flows plotted for
four depths as a function of latitude. Only OLA inversion results are shown,
RLS results are very similar. As in Figure~\ref{fig:full}, error bars are shown only
on one curve.}
\label{fig:anti}
\end{figure}

The north-south antisymmetric component of the meridional flow at four different depths is shown in Figure~\ref{fig:anti}.
 Note that the behavior is different at different 
depths and at different epochs.
The flow velocities appear to be the lowest at the
shallowest depth that we could probe. However, it is not clear if the
flow speed increases below a depth of 4.2 Mm ($r=0.994R_\odot$).

The time and latitudinal dependences of the flows is seen more clearly in
Figure~\ref{fig:merid4}. To give a symmetric appearance we have flipped
the sign of flow velocities in the southern hemisphere.
It is clear from the figure that the latitudinal dependence
of the flows is a function of time and is different at different depths. 
In particular, in the shallower layers one can see  clearly that 
the  flow velocities were smallest when the Sun was most active. 
As activity during cycle 23 rose, the flows in the higher latitudes started to
slow down first and the behavior very quickly migrated to the lower
latitudes. 
The increase in flow-speed with decreasing levels of activity is seen first in the 
high-latitude regions. This gain in speed is subsequently seen at lower
and lower latitudes.
Furthermore, the general level of the meridional flow speed in the shallow
layers  is larger during the current minimum as compared to that during the previous
minimum.
The behavior is much more complicated in the deeper layers that we can probe 
reliably, though the general pattern of the higher latitudes reacting first
appears to hold. 

The migration of the meridional flows is very similar to the butterfly diagram, as can be seen from 
Figure~\ref{fig:merid2}, where we have over-plotted the position of sunspots during cycle 
23 on the meridional flow pattern. As can be seen, there is a good match between the
positions of the sunspots and the region of low-speed meridional flows. The
concordance is not perfect, but that could be a result of the poor time and
latitude resolution
of the meridional flow data. Figure~\ref{fig:merid2}
also shows the solar zonal flow velocities at the same depth over-plotted on the meridional
flows. Again there is a match between the migrating pattern of the meridional flow
with that of the zonal flow. The zonal flow velocities were obtained using 
MDI global mode data (Antia et al.~2008). The bands of high meridional
flows coincides with those of high zonal flows.  A high speed branch, presumably
corresponding to the next solar cycle has started at high latitudes around
2004, well before the current minimum.

To study the detailed behavior of the meridional flows, we obtain a polynomial
decomposition of the flow in the manner described by Hathaway (1996)
\begin{equation}
u_y(r,\theta,t)=-\sum_i C_i(r,t)P^1_i(\cos\theta),
\label{eq:leg}
\end{equation}
where $P^1_i$ are associated Legendre polynomials of degree $i$ and order 1, and $\theta$ is the 
angle from the north pole. 
We used the standard normalization for the $P^1_i(\cos\theta)$.
Durney (1993) had suggested the presence of components beyond $i=2$.
It should be noted that Hathaway (1996)  was using only surface Doppler measurements and hence,
the  expansion-coefficients did not include any $r$ dependence.
In Eq.~(\ref{eq:leg}), and all our figures, positive $u_y(r,\theta,t)$ is the flow towards the solar
 north pole.
In this expansion, the odd-degree coefficients represent the symmetric component
of the flows and we ignore those in this work, while the even-degree polynomials are the antisymmetric component, and those
are what we concentrate on.

\begin{figure}
\epsscale{0.9}
\plotone{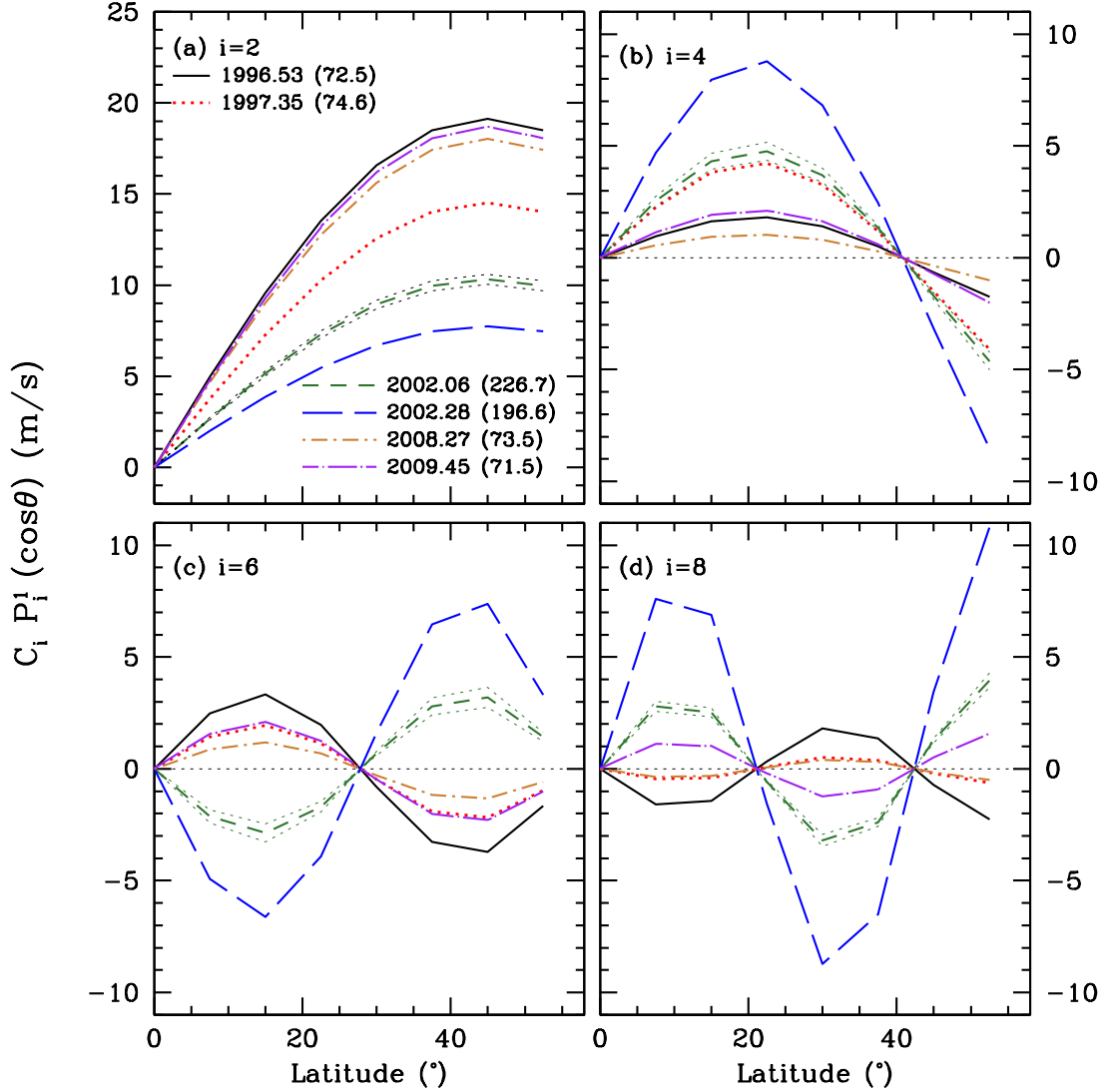}
\caption{The first 4 antisymmetric Legendre components (see Eq.~\ref{eq:leg}) of the meridional
flow at a depth of 1.4 Mm ($r=0.998R_\odot$) plotted as a function of latitude. We have plotted results
of only six sets to highlight the difference in the flows between solar minimum and
maximum and the differences between the minimum before cycle 23 and that before
cycle 24. $1\sigma$ error limits are shown for one epoch, the errors are similar
for  the others. The six sets are labeled by the central time for each data set.
The numbers in brackets in the legend indicate the 10.7 cm flux in 
SFUs.
}
\label{fig:recons}
\end{figure}

Figure~\ref{fig:recons} shows the first four even-degree Legendre components for
six epochs --- two for the beginning of cycle 23, two near the maximum and two
at the beginning of cycle 24. We plot the components for the flow at
a depth of 1.4 Mm ($r=0.998R_\odot$).
Two features stand out from the figures: first, the flow at solar maximum is not
merely slower than the flows at solar minimum, they also
have different latitudinal dependencies; 
and second, the flows at the beginning of
cycle 24 are different from the flows at the beginning of cycle 23 for comparable
levels of global solar activity. The speed of the dominant Legendre component ($i=2$) 
of the flows shows that the meridional flow was slower at the maximum than
at the minimum, though
the behavior is by no means monotonic with activity. Perhaps the more
interesting component is $i=6$, where there is a clear sign reversal of the 
flows between minimum and maximum activity. It is also clear that the higher
order components make a significant contribution to the flow.

\begin{figure}
\epsscale{0.9}
\plotone{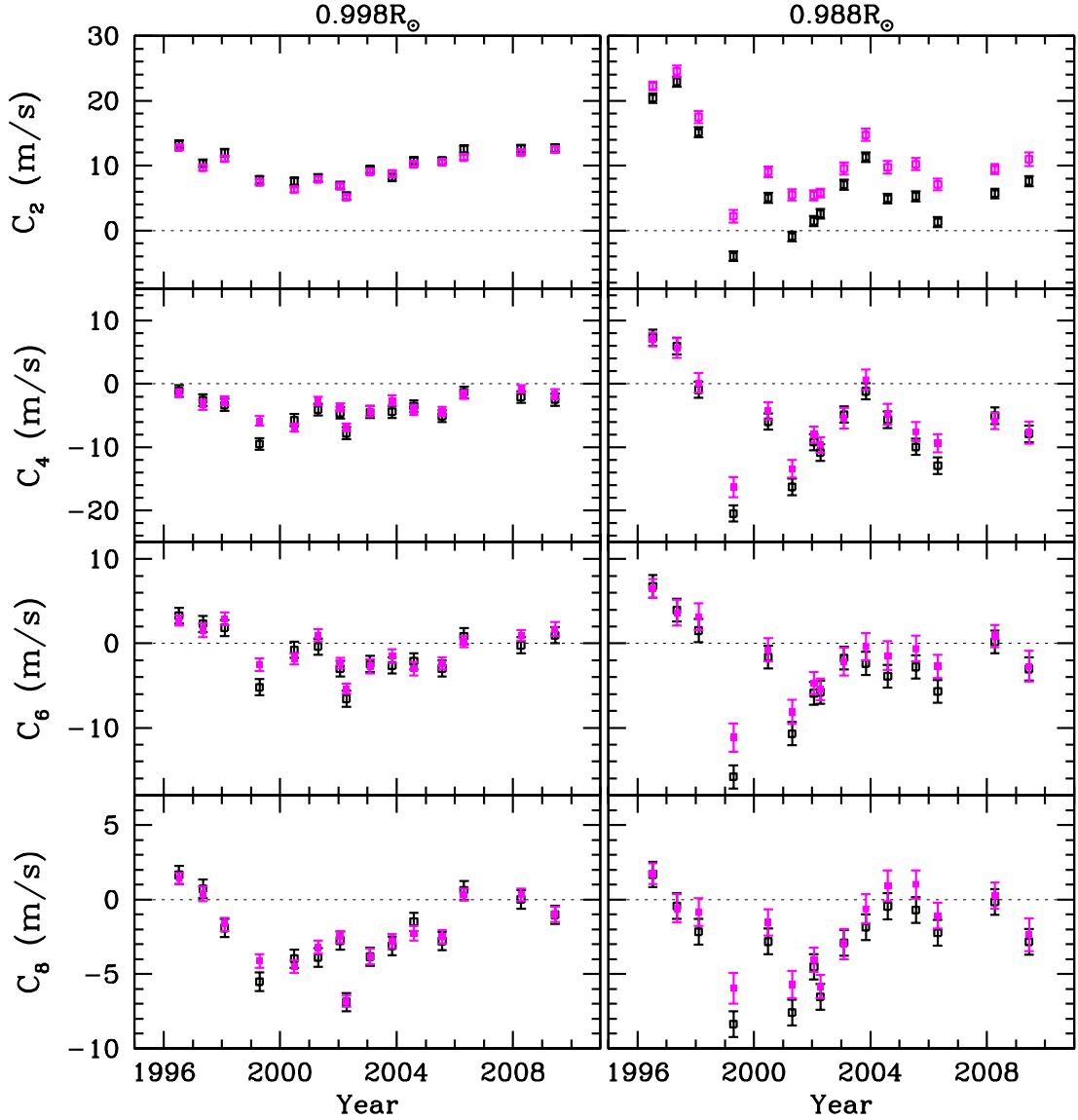}
\caption{Coefficients of the Legendre decomposition plotted as a function of time for two depths. The black points are
RLS results, the magenta (gray in the print version) are OLA results.
Note that the pattern of time dependence
is different for different coefficients and also for different radii.
}
\label{fig:coeftime}
\end{figure}

\begin{figure}
\epsscale{0.60}
\plotone{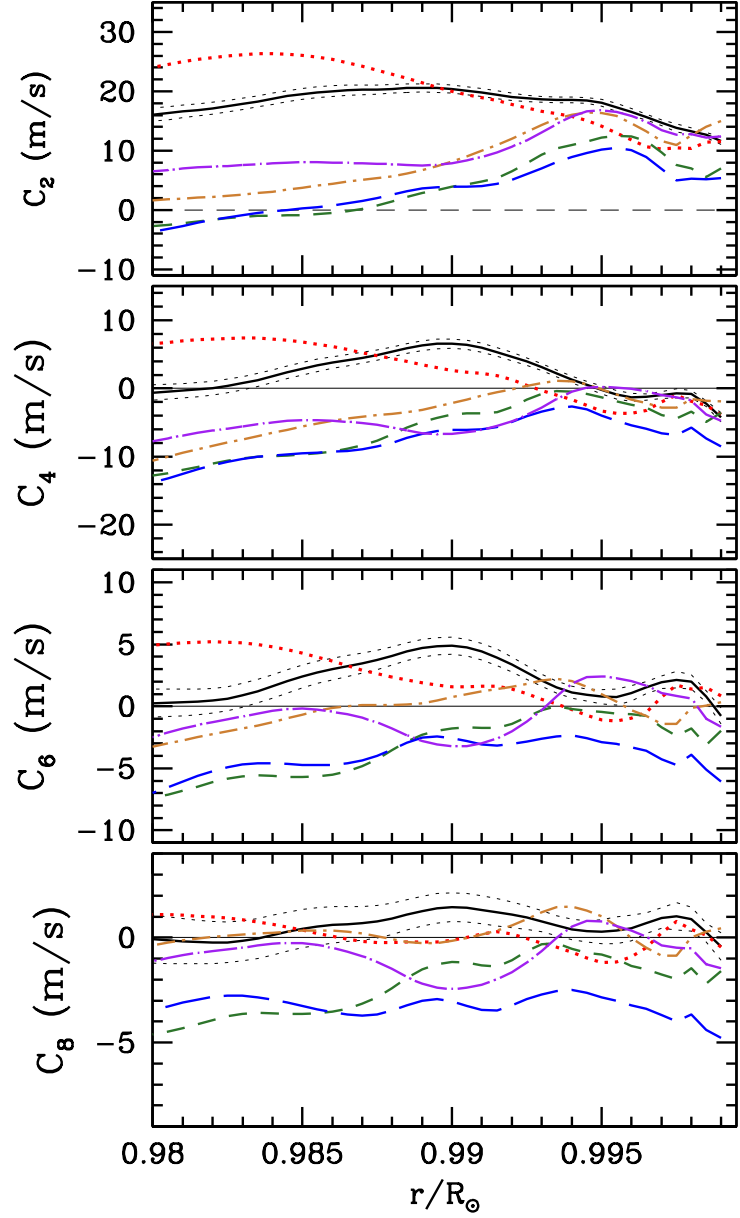}
\caption{Coefficients of the Legendre decomposition plotted as a function of depth for the six epochs
plotted in Figure~\ref{fig:recons}. The colors and line-styles in this figure are the same as those in
Figure~\ref{fig:recons}. $1\sigma$ error limits are shown only for one epoch, the errors for the
others are similar.
}
\label{fig:coefd}
\end{figure}

\begin{figure}
\epsscale{0.9}
\plotone{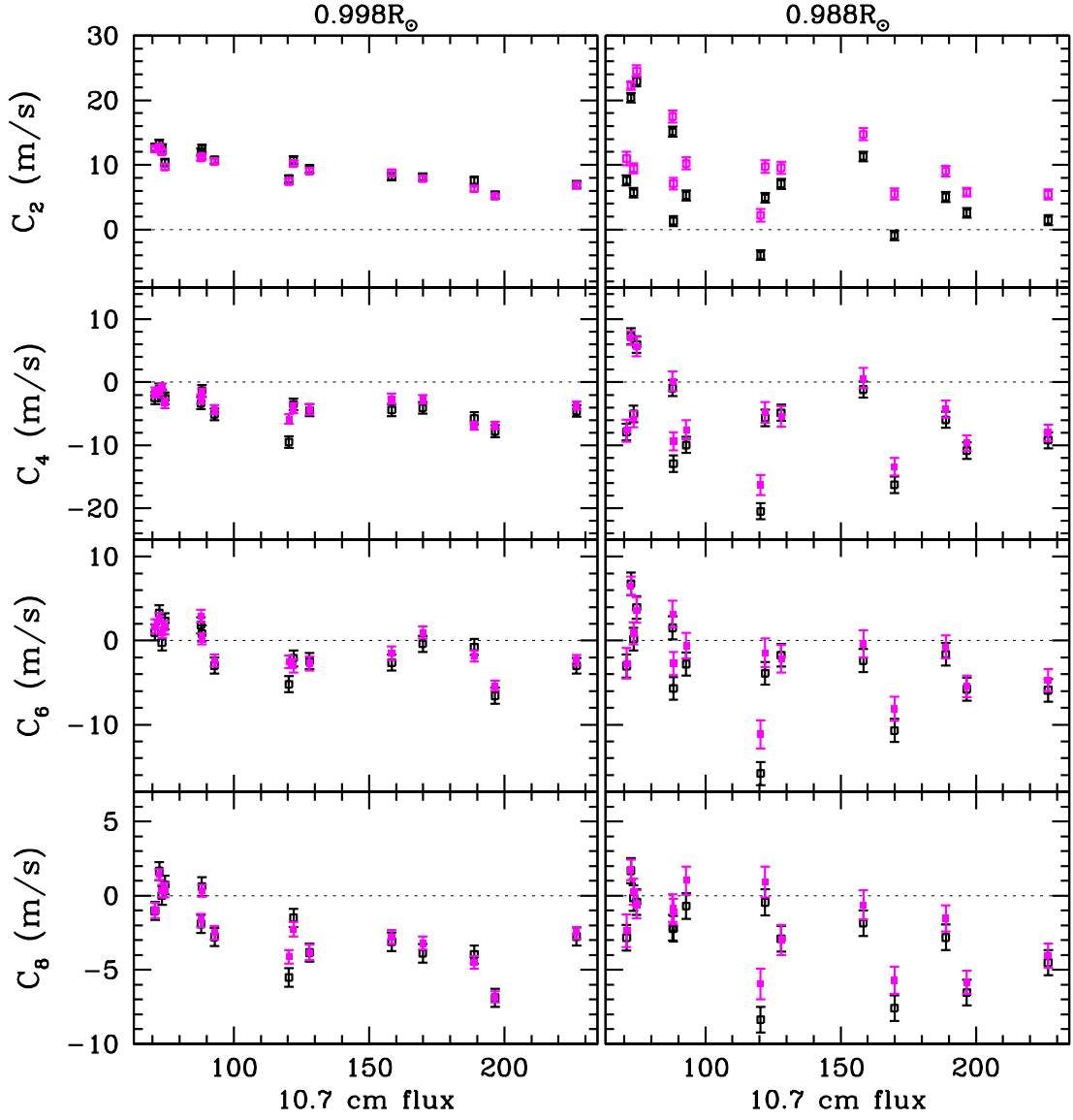}
\caption{The same as Figure~\ref{fig:coeftime}, but plotted as a function of the 10.7 cm radio
flux expressed in solar flux units.}
\label{fig:coefact}
\end{figure}

The behavior of the coefficients
at two different depths is plotted as a function of time in Figure~\ref{fig:coeftime}
and the behavior as a function of depth for six different epochs is plotted
in Figure~\ref{fig:coefd}. It is clear from the figures that (a) the coefficients
changed with time, (b) different coefficients had different time-dependences,
(c) that the coefficients were very different during the solar maximum
as compared to those at the two minima, and (d) coefficients at different depths show
different time dependence. Since each component has a different temporal
and depth dependence, the resulting pattern shows a more complex behavior.
The same data are plotted as a function of the global solar activity index (the
10.7 cm radio flux) in Figure~\ref{fig:coefact}. There appears to be a reasonable
correlation (rather an anti-correlation)
 between the $C_2$ coefficient and activity index at a depth of 1.4 Mm ($r=0.998R_\odot$), the correlation is 
less strong for the higher-order coefficients. The pattern is however, much more
confused in the deeper layers. 
While the $C_2$ coefficient is almost flat, the
higher-order coefficients show two branches. The rising phase of cycle 23 appears to
populate the lower branch,
 while the declining phase and the cycle 24 minimum populates the upper, almost
flat branch.

\begin{figure}
\epsscale{0.45}
\plotone{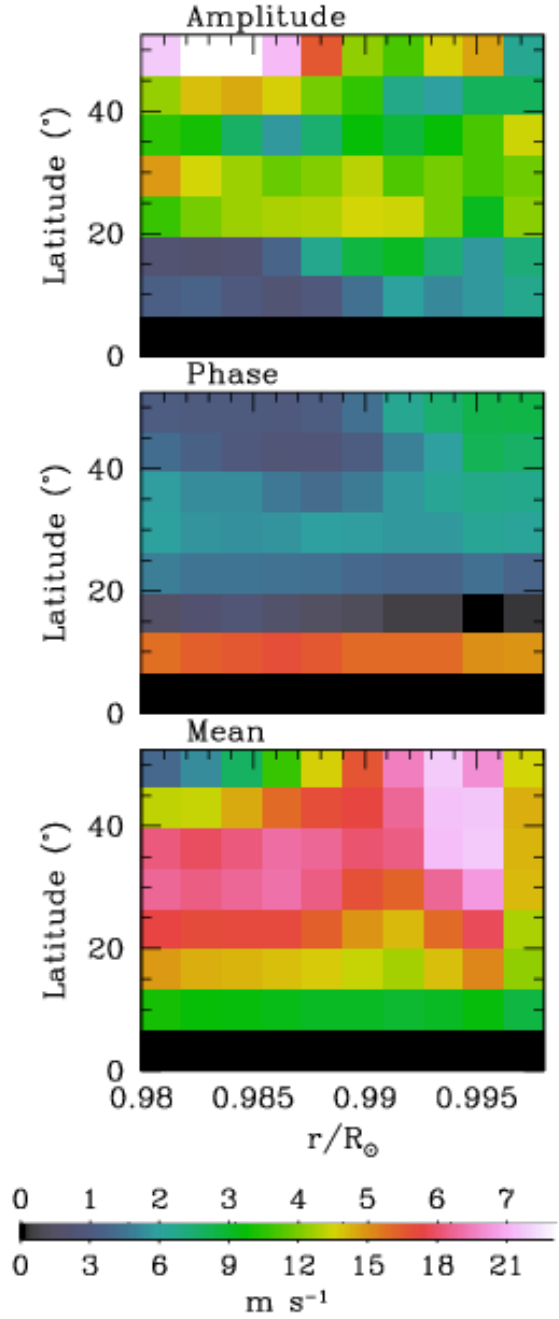}
\caption{The (temporal) mean meridional velocity as well as the amplitude
and phase of the oscillatory part of the meridional flows (see Eq.~2) are shown as a function of depth and
latitude. The lower scale on the color bar refer to the mean velocity,
while the upper one refers to the amplitude (in m s$^{-1}$) and phase
of the oscillatory component.}
\label{fig:meridt}
\end{figure}

An alternative representation of the meridional flows can be obtained by assuming that the
flow is periodic with the period of the solar cycle. Under this assumption we can
expand the flow as:
\begin{equation}
u_y(r,\theta,t)=u_a(r,\theta)+u_1(r,\theta)\sin(\omega t)+u_2(r,\theta)
\cos(\omega t),
\end{equation}
where $\omega=2\pi/T$ is the frequency of the solar cycle.
We have used $T=11.8$ years, which is the value we obtain  from
solar zonal flow
studies.
The coefficients $u_a,u_1,u_2$ are calculated at each value of
$r$ for each $\theta$. The last two coefficients thus  give the amplitude and the
phase of the time-varying component. 
The results are shown in Figure~\ref{fig:meridt}.
Since we only consider the antisymmetric component of the flow, only 
results for the northern hemisphere are shown. It can be seen that the time-averaged flow speed, $u_a$,
is maximum at intermediate latitudes. This mean velocity is small at the outermost
layer considered in this work, i.e., a depth of 1.4 Mm ($r=0.998R_\odot$), but increases steeply
with depth, and appears to decrease marginally in deeper layers. The amplitude
of the oscillatory part increases with latitude, while the phase appears to
show a complicated behavior. At the latitude of $7.5^\circ$ the phase
appears to be very large, but that is because we
have constrained the value of the phase to be between 0 and $2\pi$. 
Alternately, we can interpret the phase to be marginally negative at
this latitude. It then becomes positive  and increases with latitude. Given that
we are only showing the antisymmetric component, 7.5\dg\ is the lowest latitude for
which we have data.

\section{Discussion}
\label{sec:disc}

We have studied solar meridional flows and their variations from June 1996 to
April 2009 using the ring diagram technique applied to full-disc Dopplergrams
from MDI. This period covers the  solar cycle 23 in its entirety.
The striking feature that we find is that the flows not only vary with depth,
and latitude, but that the time-dependence of the variation  is a function of
both depth and latitude.

The meridional flows show a migrating pattern, very similar to that of the zonal flows.
This result was first seen in time-distance measurements of Beck et al.~(2002) and they 
had concluded that there is a strong correlation between the time-dependent
part of the meridional flows and the torsional oscillations (zonal flows). This was later 
confirmed using GONG data by Gonz\'alez~Hern\'andez et al.~(2010).
However, Beck et al.~(2002) only had data for the rising part of the cycle,
while Gonz\'alez~Hern\'andez et al.~(2010), using GONG data, could only study
the declining phase of cycle 23. As a result, it was not completely clear
whether the meridional flow pattern matches the zonal flow bands.
Our results confirm the migrating pattern, and also that close to the surface the migration
mimics the migration of the zonal flows.
We find that a band of faster flow at high latitudes has appeared around
2004, long before the current minimum, this is also seen with GONG data
(Gonz\'alez~Hern\'andez et al.~2010).
However, we find that although the pattern of the zonal flow is
similar from the surface down to a radius of $0.90R_\odot$ (e.g. Antia et al.~2008), 
the meridional flow pattern changes rapidly with depth.
It is possible that flow pattern at the deeper levels reflects the 
magnetic field pattern at those levels, however, the question then would be the
lack of the change in the zonal flows.
It is, however, not completely clear if the differences in the depth-dependence of the
zonal and meridional flow patterns is of a fundamental nature, or  simply
a results of the fact that ring diagram results have a higher radial resolution
near the solar surface
compared to global mode analyses used to study zonal flows.

Svanda et al.~(2008) have studied the effects of solar active regions on
meridional flows and claim that a part of the temporal variation at low latitudes
could be explained by flows around active regions. This could explain the
shifting pattern of the meridional flows, but not all of the observed  temporal variation.
However, by  masking out active regions  Gonz\'alez~Hern\'andez et al.~(2008) showed that while the
resultant flow velocity that they obtain changes in the process, there is a
second component that still shows a decreasing amplitude as magnetic activity 
increases. Thus what we are seeing is not merely a local effect of sunspot magnetic 
fields.
In addition to sunspots, there is some evidence to suggest that the variation in
the B-angle, $B_0$. with time also may cause apparent temporal variation
in meridional flows (Zaatri et al.~2006). This has been argued to be the
reason why some results showed presence of counter cells in layers below
the surface (Haber et al.~2002; Gonz\'alez Hern\'andez et al.~2006).
However, in this work we have used only the antisymmetric component of
the meridional flow, which minimizes the effect of $B_0$ angle
variation. Thus we do not expect our results to be significantly
affected by variation in $B_0$. The fact that we see a consistent
variation in the meridional-flow velocity supports this conclusion.

In order to get a better understanding of the flow pattern, we decompose
the latitudinal dependence in terms of associated Legendre polynomials.
This decomposition of meridional flows shows that higher degree components
also make a significant contribution. These components do not have the same
sign at all latitudes in a given hemisphere, and hence, they give rise to a complex pattern of variation
in the resulting flows.
These components, in particular, the $P^1_6$ and $P^1_8$ components, change
sign with time and are reversed during the solar
maximum with respect to the flow during the solar minimum.
While the higher order components
show a time-variation, the changes are not clearly correlated to solar
activity. This is probably the result of the fact that 
global activity indices are obtained by averaging over the disc while
the higher components of flows have a complex latitudinal dependence.

The dominant component of meridional flow ($P_2^1(\cos\theta)$) shows a
clear variation with time that is anti-correlated with global solar activity. Chou \& Dai
(2001) also found similar results. Basu \& Antia (2003) had also found similar variations but the
result was not very clear with the limited data available at that time.  
The extended minimum before cycle 24 clearly shows up in the variation
of coefficient $C_2$ in the shallower layers. There is no sign until the last
data set (June 2009) of $C_2$ starting to decrease. Our results indicate that the magnitude
of this component is similar between the two minima of cycles 23 and 24.
Our results are qualitatively similar to those found by Hathaway \& Rightmire (2010) from
magnetic-field observations made by MDI, though the amplitude they find is
consistently lower than what we find after correcting for the difference
in normalization. We get $C_2$ values of $12.8\pm 0.4$ m s$^{-1}$, $6.9 \pm 0.4$  m s$^{-1}$
and $12.5 \pm 0.5$ m s$^{-1}$ respectively, for mid-1996, early 2002 and mid 2009
at a depth of 1.4 Mm ($r=0.998R_\odot$).
It should be noted that  Hathaway \& Rightmire (2010) report the 
amplitude of the $2\sin\theta\cos\theta$ term
while we use $P_2^1(\cos\theta)=-3\sin\theta\cos\theta$. Thus our $C_2$
should be multiplied by 1.5 before comparing with their values.
Hathaway \& Rightmire (2010) also find the coefficient $C_2$
to be larger during the current minimum as compared to that during
the previous minimum. 
We however, do not find a significant difference  at the lowest activity levels
that we have access to, however $C_2$ during slightly higher activity levels
(albeit still low activity) are somewhat different during the
minimum of cycle 24 compared with those of cycle 23. The differences
are clearly seen in the coefficients of the higher order components in Legendre
decomposition.
As can be seen from Table~1, the activity level during the first set in
our list is the same as that during the last set.
 Unfortunately, we do
not have data for December 2008, when solar activity was lowest.
It is quite possible that the difference between our 
results and those of Hathaway \& Rightmire (2010) is because we are looking at 
different depths, or because of the different observables and techniques
used in the two studies. Our results show a rapid variation of
flow amplitudes with depth and it is  not clear what depth is 
probed by the magnetic features used by Hathaway \& Rightmire (2010).

\begin{figure}
\epsscale{0.9}
\plotone{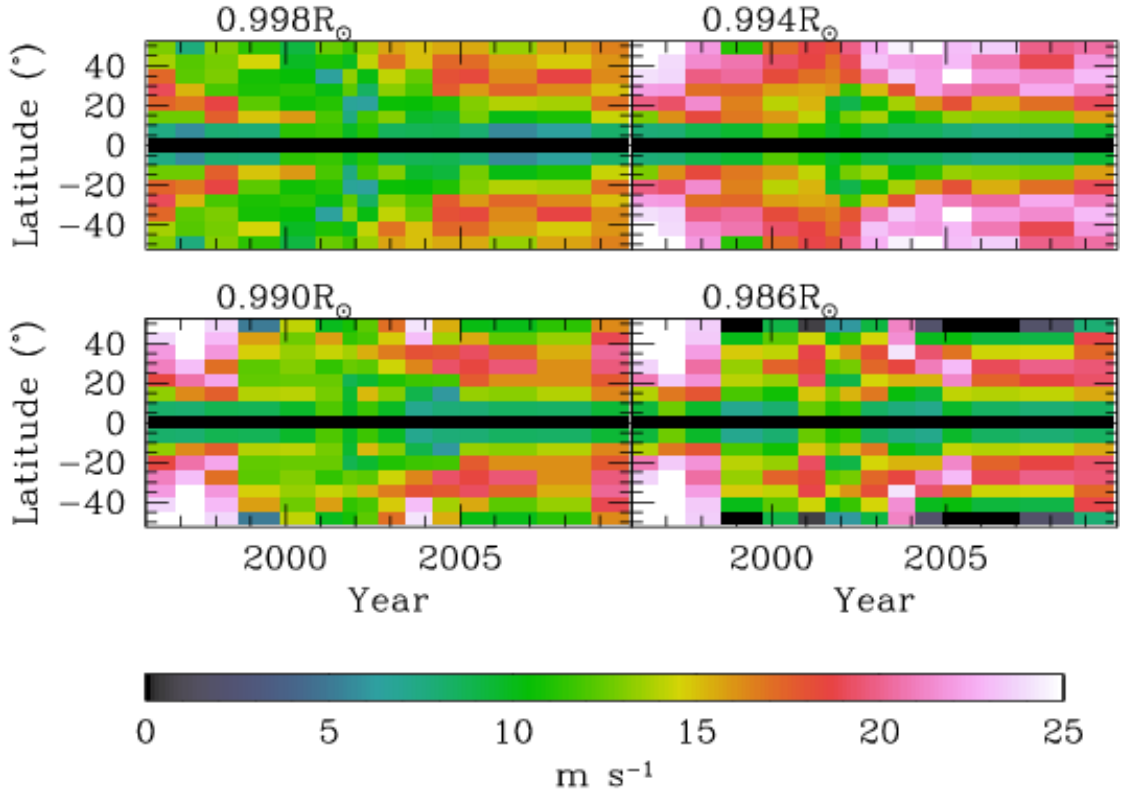}
\caption{The speed of the antisymmetric component of the meridional flows plotted as a function of 
latitude and time at four depths.
To make the plot symmetric, the sign of flow has been reversed in the southern
hemisphere. Thus positive velocity points towards the poles in both
hemispheres.
Each data point is assumed to represent the time interval between the mid-points
with neighboring sets. 
The actual time
covered by each set (which is typically one Carrington rotation) is usually much smaller than the interval
between the sets.
}
\label{fig:merid4}
\end{figure}

\begin{figure}
\epsscale{0.6}
\plotone{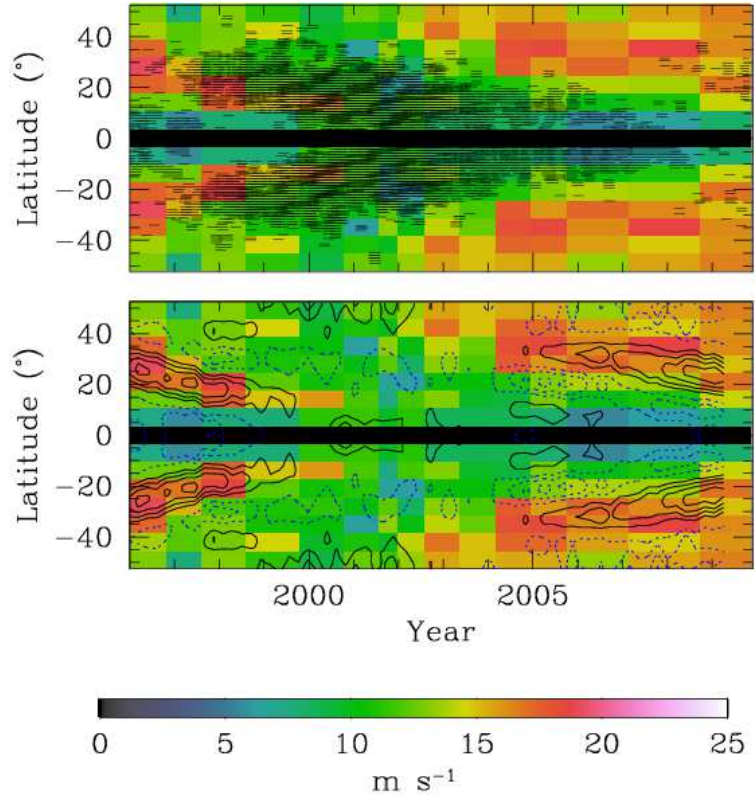}
\caption{The antisymmetric component of meridional flow plotted as a function of time and latitude at a depth of 1.4 Mm ($r=0.998R_\odot$) (cf., Fig.~\ref{fig:merid4})
over-plotted with the position of sunspots (upper panel) and contours showing zonal flow velocities
(lower panel). The solid contours are for prograde velocities, dashed for retrograde. In the lower panel contours are
spaced at intervals of 1m s$^{-1}$ and the zero-velocity contour has not been plotted.
}
\label{fig:merid2}
\end{figure}

The strength of the solar meridional flow is believed to play an important role
in transporting magnetic flux. Devore et al.~(1984) show that the meridional
flows play a significant role in magnetic field distribution at low activity.
A lower meridional flow speed tends to give stronger polar fields (Sheeley et al.
1989).
Our results (see Figure~\ref{fig:merid4}) would imply a
weaker polar field during the minimum of cycle 24.
That is indeed the case. Wang et al.~(2009) find
that the  polar fields during the
minimum of cycle 24 were about 40\% weeker than that during the earlier minimum.

As mentioned earlier, we find that the meridional flow speed increases with depth.
The increase in meridional flow amplitudes with depth in the near-surface
layers is seen in {\it ad hoc} models of the solar  meridional flow (e.g., Roth et al.~2002;
Chatterjee \& Antia 2009). In these models the increase can be attributed
to the boundary condition of radial velocity vanishing at the top boundary
and the small density scale heights in the subsurface layers. In these
models the meridional velocity has a maximum close to the surface, around
a depth of 3.5 Mm ($r=0.995R_\odot$), and then
decreases becoming significantly less below a depth of 14 Mm ($r=0.98R_\odot$).
Since these features are essentially a result of the continuity equation, similar
variations may be seen in more realistic profiles or in the Sun.
On the other hand, Spruit (2003) has proposed a model for torsional
oscillations as a geostrophic flow due to lower subsurface temperatures in
active regions. This model also gives rise to a meridional flow. However, it
predicts that the meridional flow amplitude would decrease with
depth, which is opposite to what we find. If this trend is confirmed, such
models will be ruled out.

Our results are for very shallow layers of the Sun; we can only probe the outer
2\% of solar radius with any reliability.
There would be considerable scientific return if the
analysis could be extended to deeper layers not covered in this work.
Gonz\'alez Hern\'andez et al~(2006) have tried getting results for deeper
layers of the Sun using ring diagrams of regions of size 32\dg\ (in contrast, this and other
ring diagram work use 16\dg\ regions) and find they can cover the outer 5\% of the Sun. However, this
technique has the drawback that it necessarily degrades the latitudinal resolution. Perhaps a 
combination of rings-diagrams of different sizes is required to get a good coverage. Time-distance
helioseismology could, in principle, give us information of flows in the deeper layer, but
again there would be a tradeoff in terms of latitudinal resolution. Data from the Helioseismic and Magnetic Imager (HMI)
on board the recently launched Solar Dynamics Observatory will undoubtedly provide us 
with a better picture of solar meridional flows, though
analyzing meridional flows in the deeper layers will remain a problem. The data, however, will 
allow us to study cycle-to-cycle variations
of the flows.  The higher resolution of HMI will also
allow us to extend helioseismic meridional flow studies to higher latitudes.

\acknowledgments
This work utilizes data from the Solar Oscillations Investigation/ Michelson 
Doppler Imager (SOI/MDI) on the Solar and Heliospheric Observatory (SOHO). 
SOHO is a project of international cooperation between ESA and NASA. 
SB acknowledges support from NSF grants ATM 0348837 and ATM 0737770 
and NASA grants NNG06D13C and NXX10AE60G.

\newpage

\begin{deluxetable}{ccl}
\tablecaption{\em Data sets analyzed}
\tablehead{\colhead{Carrington}& \colhead{Dates} &\colhead{10.7 cm Flux}\\
\colhead{Rotation}&&\quad(SFU)$^1$\\}
\startdata
1910 & 1996:06:01 -- 1996:06:28 & $\phantom{0}71.7\pm 0.3$\\
1922 & 1997:04:24 -- 1997:05:22 & $\phantom{0}74.6\pm 0.6$\\
1932 & 1998:01:22 -- 1998:02:18 & $\phantom{0}87.9\pm 1.5$\\
1948 & 1999:04:04 -- 1999:05:01 & $120.4\pm 2.5$\\
1964 & 2000:06:13 -- 2000:07:10 & $188.9\pm 3.8$\\
1975 & 2001:04:09 -- 2001:05:06 & $169.9\pm 4.6$\\
1985 & 2002:01:11 -- 2002:02:02 & $226.7\pm 3.1$\\
1988 & 2002:03:30 -- 2002:04:26 & $196.6\pm 3.1$\\
1999 & 2003:01:25 -- 2003:02:10 & $128.0\pm 2.1$\\
2009 & 2003:10:22 -- 2003:11:17 & $158.4\pm 13.7$\\
2019 & 2004:07:22 -- 2004:08:18 & $122.1\pm 5.2$\\
2032 & 2005:07:11 --  2005:08:08 & $\phantom{0}92.9 \pm 2.4$\\
2042 & 2006:04:10 -- 2006:05:08 & $\phantom{0}88.2\pm 1.6$\\
2068-69 & 2008:03:24 -- 2008:04:27 & $\phantom{0}73.5\pm 0.9$\\
2083-84 & 2009:05:20 -- 2009:06:16 & $\phantom{0}71.5\pm 0.3$\\
\enddata
\tablenotetext{1}{ The `errors' are the standard deviation of the daily
10.7 cm flux during the interval covered by each data set.}
\label{tab:data}
\end{deluxetable}
\newpage

\end{document}